\documentclass[epj]{svjour}
\usepackage{graphics}
\usepackage{amsmath}
\begin{document}
\title{Electroweak Physics at the Tevatron and LHC: Theoretical Status
and Perspectives}
\author{Ulrich Baur
}                     
\offprints{}          
\institute{Physics Department, State University of New York at Buffalo,
Buffalo, NY 14260, USA}
\date{Received: date / Revised version: date}
%
\abstract{
I review the status of theoretical calculations relevant for
electroweak physics at the Tevatron and LHC and discuss future
directions. I also give a brief overview of current electroweak data and
discuss future expectations.
\PACS{
      {PACS-key}{describing text of that key}   \and
      {PACS-key}{describing text of that key}
     } 
} 
\maketitle
\section{Introduction}
\label{intro}
Electroweak measurements are a very important part of the physics program of
the Tevatron and the LHC. Of particular interest are the search for the
Higgs boson and the determination of its properties, and the measurement
of electroweak precision observables, in particular the measurement of 

\begin{itemize}

\item the $W$ mass, $M_W$, and $W$ width, $\Gamma_W$,

\item the effective weak mixing angle, $\sin^2\theta_{eff}$, and the
forward -- backward asymmetry, $A_{FB}$,

\item the $W$ and $Z$ boson cross sections, $\sigma(W)$ and $\sigma(Z)$,
and their ratio, $R_{W/Z}$,

\item the $W$ forward backward charge asymmetry, $A(\eta_e)$,

\item the $\ell^+\ell^-$ ($\ell\nu$) differential cross sections above
the $Z$ ($W$) peak,

\item and di-boson ($W\gamma$, $Z\gamma$, $WW$, $WZ$ and $ZZ$)
production.

\end{itemize}

In the following I discuss the physics interest in these
measurements, give a brief overview of the current experimental status
and what to expect in the future (for more details see
Refs.~\cite{serban} --~\cite{goshaw}), and discuss
the current status and the prospects of the relevant theoretical
calculations.

\section{Weak Boson Physics}
\label{sec:1}

\subsection{Measurement of the $\boldsymbol{W}$ mass}
\label{sec:1.1}

The one-loop corrections to $M_W$ depend quadratically on the top quark
mass, $m_t$, and logarithmically on the Higgs boson mass, $m_H$. Precise
measurements of $M_W$ and $m_t$ thus make it possible to extract
information on $m_H$. 

In Run~I of the Tevatron, the $W$ mass has been
measured to $M_W=80.456\pm 0.059$~GeV~\cite{Abazov:2003sv}. The
preliminary value of the $W$ mass from LEP2 is $M_W=80.392\pm
0.039$~GeV~\cite{grun}. When combined with the current world average of
the top quark mass, $m_t=172.7\pm 2.9$~GeV~\cite{unknown:2005cc}, this
yields $m_H<219$~GeV at 95\% CL~\cite{grun} for a Standard Model
(SM) Higgs boson. 

\begin{figure}
\begin{center}
\resizebox{0.45\textwidth}{!}{
  \includegraphics{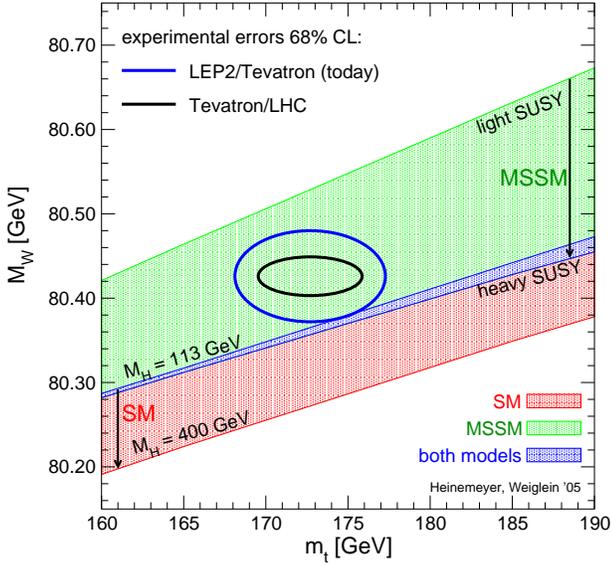}}
\caption[]{Constraints on $M_W$ and $m_t$ from LEP2 and Tevatron data,
and expectations from the LHC, compared with the predictions of the SM
and the MSSM.}
\label{fig:two}       
\end{center}
\end{figure}

In Run~II, one hopes to achieve a precision of $\delta M_W=40$~MeV per
lepton channel and experiment for an integrated luminosity of
2~fb$^{-1}$~\cite{Brock:1999ep}, while the LHC may be able to reach a
precision of $\delta M_W\approx 10$~MeV using the $W/Z$ transverse mass
ratio and $W\to\mu\nu$ decays~\cite{schmidt}. The present constraints on
$M_W$ and $m_t$ from LEP2 and Tevatron data, and the results expected
from measurements at the LHC, are compared with theoretical predictions
in Fig.~\ref{fig:two}. Present data clearly favor a light SM Higgs boson,
and are also in very good agreement with predictions of the minimal
supersymmetric standard model (MSSM)~\cite{schweiglein}.

To ensure that the $M_W$ and $m_t$ measurements contribute equally to
the uncertainty in a $\chi^2$ test, the precision on the top quark mass
and the $W$ mass should satisfy the relation~\cite{Haywood:1999qg}
\begin{equation}
\delta M_W\approx 7\times 10^{-3}\,\delta m_t.
\end{equation}
Since one expects to measure the top quark mass with a precision of
$\delta m_t=1-2$~GeV at the LHC~\cite{Beneke:2000hk}, one needs to
determine the $W$ mass with a precision of about $\delta M_W\approx
10$~MeV so that it does not become the dominant uncertainty in the
estimate of $m_H$. Accurate 
theoretical predictions for $W$ production are absolutely essential in
order to measure the $W$ mass with a precision of 10~MeV.

\subsection{$\boldsymbol{\sin^2\theta_{eff}}$}

Constraints on $m_H$ can also be derived from the top quark mass and the
effective weak mixing angle. At LEP, the effective weak mixing angle has
been measured to $\sin^2\theta_{eff}=0.23153\pm
0.00016$~\cite{physrep}. This will be difficult to improve at the
Tevatron or LHC. From a measurement of the forward -- backward
asymmetry, $A_{FB}$, at the Tevatron one expects to reach a precision of
$\delta\sin^2\theta_{eff}=0.0006$ per lepton channel and experiment for
an integrated luminosity of 
10~fb$^{-1}$~\cite{Brock:1999ep}. At the LHC, with 100~fb$^{-1}$, one
hopes to achieve $\delta\sin^2\theta_{eff}=0.00014$ using forward
electrons in a measurement of $A_{FB}$ in $Z\to e^+e^-$
events~\cite{atlnote}. 

\subsection{The Weak Boson Cross Sections and the $\boldsymbol{W/Z}$
Cross Section Ratio} 

In the past, the measurement of the $W$ and $Z$ boson cross sections has
provided a test of perturbative QCD. With the large data sets of Run~II
and the LHC, non-statistical uncertainties, in particular the luminosity
error, become limiting factors. This is illustrated by the recent D\O\
measurements of the $W$ and $Z$ production cross sections~\cite{d0cross}
\begin{eqnarray*}
\sigma(W)\cdot B(W\to e\nu) & = & 2865.2\pm 8.3{\rm (stat)} \pm 62.8{\rm
(sys)} \\
& & \pm 40.4{\rm (pdf)}\pm 186.2{\rm (lumi)~pb,}\\
\sigma(Z)\cdot B(Z\to e^+e^-) & = & 264.9\pm 3.9{\rm (stat)} 
\pm 8.5{\rm (sys)} \\
& & \pm 5.1{\rm (pdf)}\pm 17.2{\rm (lumi)~pb.}
\end{eqnarray*}
Provided that the $W$ and $Z$ cross sections are accurately predicted by
theory, and the PDF uncertainties can be controlled, $\sigma(W)$ and
$\sigma(Z)$ can be used as luminosity monitors.

The cross section ratio
\begin{equation}
R_{W/Z}={\sigma(p\bar p\to W\to\ell\nu X)\over\sigma(p\bar p\to Z
\to\ell^+\ell^- X)}~,
\end{equation}
together with the theoretical prediction for the ratio of the total $W$
and $Z$ production cross sections, the LEP measurement of 
the branching ratio
$B(Z\to\ell^+\ell^-)$, and the SM prediction for the $W\to\ell\nu$ decay
width, can be used for an indirect determination of $\Gamma_W$. A recent
CDF measurement, $\Gamma_W=2079\pm 41$~MeV~\cite{Abulencia:2005ix}, is
in good agreement with the 
SM prediction $\Gamma_W^{SM}=2092\pm 3$~MeV~\cite{Hagiwara:2002fs}.

\subsection{Direct Measurement of $\boldsymbol{\Gamma_W}$}
\label{sec:width}

The width of the $W$ boson can also be measured directly from the tail
of the $W\to\ell\nu$ ($\ell=e,\,\mu$) transverse mass ($M_T$)
distribution. Unlike the extraction of $\Gamma_W$ from $R_{W/Z}$, the
measurement from the tail of the $M_T$ distribution does not depend on
theoretical assumptions; however, the method is currently not as precise
as the measurement using $R_{W/Z}$. This is illustrated by the recent
combined Tevatron result, $\Gamma_W=2078\pm 62({\rm stat})\pm 60({\rm
syst})$~MeV~\cite{d0width}. For 2~fb$^{-1}$ one expects the direct
measurement of $\Gamma_W$ to
improve to $\delta\Gamma_W=50$~MeV per lepton channel and
experiment~\cite{Amidei:1996dt}. 

\subsection{The $\boldsymbol{W}$ charge asymmetry}

\begin{figure}
\begin{center}
\resizebox{0.45\textwidth}{!}{
  \includegraphics{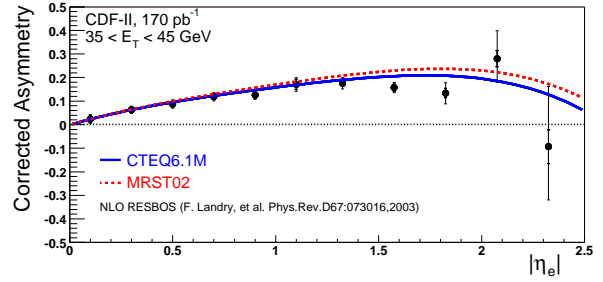}}
\caption[]{The $W$ charge asymmetry as a function of $\eta_e$ for
electrons with $35~{\rm GeV}<E_T<45$~GeV and two different PDF
parametrizations. } 
\label{fig:three}       
\end{center}
\end{figure}

Another important electroweak measurement is that of the $W$ charge
asymmetry, 
\begin{equation}
A(\eta_e)={d\sigma(e^+)/d\eta_e-d\sigma(e^-)/d\eta_e\over 
d\sigma(e^+)/d\eta_e+d\sigma(e^-)/d\eta_e}~,
\end{equation}
where $\eta_e$ is the rapidity of the electron in $W\to
e\nu$. $A(\eta_e)$ is sensitive to the $u$- and $d$-quark components of
the PDFs, especially at large values of $\eta_e$ and the electron
transverse energy, $E_T$. Results from CDF for 170~pb$^{-1}$ of Run~II
data~\cite{Acosta:2005ud} are shown in Fig.~\ref{fig:three}.

\subsection{Search for New Physics in Drell-Yan Production}

Many models of new physics predict new charged ($W'$) or neutral ($Z'$)
gauge bosons. One can search for these particles in the high
$\ell^+\ell^-$ ($\ell\nu$) invariant (transverse) mass
region. Information on the couplings of a $Z'$ boson can be obtained
from the forward -- backward asymmetry, $A_{FB}$, at large di-lepton
masses. 
Present D\O\ data (200~pb$^{-1}$) require that $m_{Z'}>780$~GeV at 95\%
CL for a SM-like $Z'$ boson~\cite{d0zprime}.

At the LHC, one can discover a $Z'$ boson with $m_{Z'}=4-5$~TeV for
300~fb$^{-1}$ and one will be able to severely constrain the couplings of
the new vector boson~\cite{cmszprime}.

\subsection{Theory of Single Weak Boson Production}

The precision foreseen for electroweak measurements in Run~II and at the
LHC has to be matched by precise theoretical predictions, ie. QCD and
electroweak (EW) radiative corrections have to be under control. 

The QCD corrections to the total $W$ and $Z$ boson cross sections at the
next-to-next-to-leading (NNLO) level have been known for more than a
decade~\cite{vanNeerven:1991gh}. Recently, the rapidity distribution of
the $Z$ boson has been calculated at NNLO, showing a dramatically
reduced dependence of the differential cross section on the unphysical
renormalization and factorization scales compared with the NLO
prediction~\cite{Anastasiou:2003yy}. Calculations of the resummed QCD 
corrections to predict the transverse momentum ($q_T$) distributions of
the $W$ and $Z$ bosons are also available~\cite{Balazs:1995nz}. The precise
shape of the weak boson $q_T$ distribution for small transverse momenta,
however, is still uncertain, in particular at the
LHC~\cite{Berge:2004nt}. 

With the the uncertainty from unknown higher order QCD corrections
approaching the 1\% level~\cite{Anastasiou:2003yy}, EW
radiative corrections to weak boson production become
important. EW corrections may also be enhanced by collinear
logarithms near the $W$ and $Z$ resonances, and by Sudakov logarithms at
large $\ell^+\ell^-$ and $\ell\nu$ invariant masses. A consistent
calculation of EW radiative corrections requires parton distribution
functions (PDFs) which take into account QED corrections. Such PDFs
exist now~\cite{Martin:2004dh}.

There has been significant progress in the calculation of the EW
radiative corrections to $W$ and $Z$ boson production in the past few
years. Calculations of the full ${\cal O}(\alpha)$ EW corrections are
available now~\cite{Baur:2001ze,Dittmaier:2001ay}. 

\begin{figure}
\begin{center}
\resizebox{0.45\textwidth}{!}{
  \includegraphics{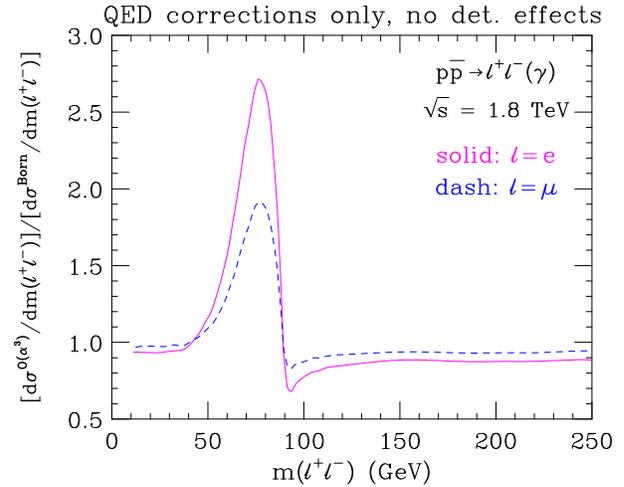}}
\caption[]{Ratio of the \protect{${\cal O}(\alpha^3)$} and lowest order
differential cross sections as a function of the di-lepton invariant
mass for $p\bar p\to\ell^+\ell^-(\gamma)$ at 
$\protect{\sqrt{s}=1.8}$~TeV. }  
\label{fig:five}       
\end{center}
\end{figure}

The main effect of the EW corrections in the vicinity of the $W$ and $Z$
resonances is that they shift the $W$ and $Z$ masses extracted from
data. The magnitude of the shift is about 50~MeV (150~MeV) for $W\to
e\nu$ ($W\to\mu\nu$). Since both leptons can radiate photons, the shifts
are about twice as large in $Z$ events. The shift is mostly caused by final
state photon radiation which is 
enhanced due to collinear logarithms of the form
$(\alpha/\pi)$ $\log(M_{W,Z}^2/m^2_\ell)$, where $m_\ell$ is the mass of
the charged lepton in $W\to\ell\nu$ or $Z\to\ell^+\ell^-$. Final state
photon radiation distorts the shape of the Breit-Wigner resonance by
reducing the peak cross section by $20-30\%$, and (significantly)
enhancing the cross section below the peak. This is shown in
Fig.~\ref{fig:five} for the $Z$
case with only QED corrections taken into account~\cite{Baur:1997wa}. 

\begin{figure}
\begin{center}
\resizebox{0.45\textwidth}{!}{
  \includegraphics{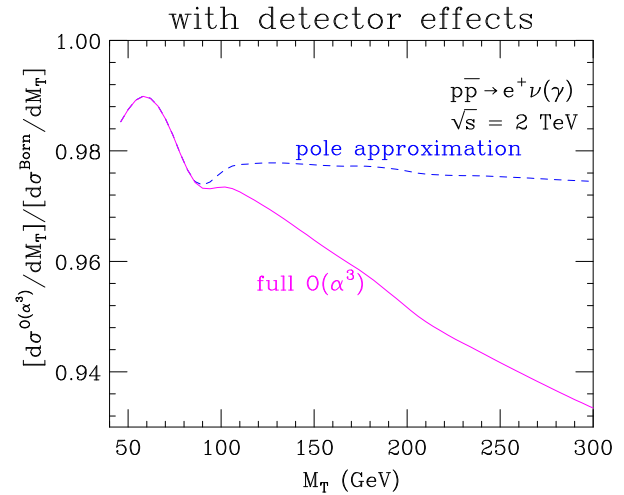}}
\caption[]{The ratio $[d\sigma^{{\cal O}(\alpha^3)}/dM_T]/[d\sigma^{
Born}/dM_T]$ as a function of the transverse mass for $p\bar p\to
e^+\nu_e(\gamma)$ at $\sqrt{s}=2$~TeV.  }  
\label{fig:six}       
\end{center}
\end{figure}

Above the $W/Z$ peak, the purely weak corrections become increasingly
important, due to Sudakov-like logarithms of the form $(\alpha/ \pi)
\log^2(\hat s/M_{W,Z}^2)$, where $\hat s$ is the squared invariant
mass of the $\ell^+\ell^-$ or $\ell\nu$ system. This is
shown in Fig.~\ref{fig:six} for the $e\nu$ transverse mass. The solid
line shows the ratio $[d\sigma^{{\cal O}(\alpha^3)}/dM_T]/[d\sigma^{
Born}/dM_T]$ taking into account the complete ${\cal O}(\alpha)$ EW
corrections. The dashed line shows the ratio in the pole
approximation~\cite{Baur:2001ze,Dittmaier:2001ay} where the $WZ$ box diagrams
responsible for the Sudakov logarithms are absent. 

Since the logarithmic terms from the $WZ$ box diagrams change the slope
of the $M_T$ distribution, they shift the $W$ width extracted from the
tail of the $M_T$ distribution. This shift, $\delta\Gamma_W\approx
-7$~MeV~\cite{Baur:2001ze}, while not large, cannot be neglected if the
Run~II goal (see Sec.~\ref{sec:width}) should be met. 

At the LHC, it will be possible to probe di-lepton invariant and
$\ell\nu$ transverse masses of several TeV. In this region, the Sudakov
logarithmic terms grow so large that they have to be resummed. Although
the resummation of 
electroweak Sudakov-like logarithms in general four fermion electroweak
processes has been discussed in the literature~\cite{cc,penin}, 
a calculation of $\ell\nu$ 
production in hadronic collisions which includes resummation of electroweak
logarithms has not been carried out yet. 

Since final state photon radiation causes a significant shift in $M_W$
and $M_Z$, one has to worry about multiple (final) state photon radiation
in weak boson production. Two photon radiation is known to considerably
change the shape of the dilepton and $\ell\nu$ transverse mass
distributions~\cite{Baur:1999hm}. Recently, there have been several
calculations of multi-photon radiation in
$W$~\cite{Placzek:2003zg,CarloniCalame:2003ck} and $Z$
decays~\cite{CarloniCalame:2005vc}. The shift in the weak boson masses
caused by multiple photon radiation is found to be about 10\% of the
shift originating from one-photon
emission~\cite{CarloniCalame:2003ck,CarloniCalame:2005vc}. For the muon
final state, where the shift in the weak boson masses is particularly
large, this is a non-negligible effect. 

The experimental precision which can be achieved for $M_W$ strongly
depends on how well the transverse momentum distribution of the $W$ is
known. Knowledge of the $W$ $q_T$ distribution determines the missing
transverse energy ($E\!\llap/_T$) resolution in $W$ events. The
$E\!\llap/_T$ resolution determines how ``sharp'' the edge in the $M_T$
distribution at $M_T\approx M_W$ is, which in turn determines how well
$M_W$ can be measured. To constrain the $W$ $q_T$ distribution, one uses
data on the transverse momentum distribution of the $Z$ boson, together
with a theoretical prediction for the ratio
$[d\sigma(W)/dq_T(W)]/[d\sigma(Z)/dq_T(Z)]$. For the $W$ mass
measurement one thus needs a calculation which includes both the resummed
QCD corrections, the full ${\cal O}(\alpha)$ EW corrections, and effects
from multiple photon radiation. A first step towards this lofty goal has
been taken in Ref.~\cite{Cao:2004yy}, where final state photon
radiation was added to a calculation of $W$ boson production which 
includes resummed QCD corrections. 

\section{Di-boson Production}

\subsection{Experimental Results}

Di-boson production makes it possible to probe the $WW\gamma$, $WWZ$,
$Z\gamma\gamma$, $ZZ\gamma$ and $ZZZ$ self-couplings (TGCs). For details
on these couplings and recent TGC measurements at the Tevatron see
Ref.~\cite{goshaw}. In the following I concentrate on the $WW\gamma$ and
$WWZ$ couplings and briefly summarize recent experimental results for
these couplings. 

The most general $WWV$ ($V=\gamma,\,Z$) vertex consistent with Lorentz
invariance and electromagnetic gauge invariance has seven free
parameters~\cite{Hagiwara:1986vm}. Assuming $C$ and $P$ conservation,
five independent couplings, $g_1^Z$, $\kappa_V$ and $\lambda_V$,
remain. Requiring SU(2) invariance as well, $\lambda_Z=\lambda_\gamma$ and
$\kappa_Z=g_1^Z-(\kappa_\gamma-1)\tan^2\theta_W$, where $\theta_W$ is
the weak mixing angle~\cite{Hagiwara:1992eh}, and one is left with three
independent couplings. In the SM, at tree level,
$g_1^Z=\kappa_V=1$ and $\lambda_V=0$. In order to avoid that $S$-matrix
unitarity is violated, deviations of the TGCs from their SM values have
to be momentum dependent form factors which depend on the form factor
scale, $\Lambda$~\cite{Baur:1987mt}. The form factor scale is related to
the scale of the new physics which causes non-SM TGCs.

The $WWV$ couplings can be measured in $e^+e^-\to W^+W^-$, and in
$W\gamma$, $WZ$ and $WW$ pair production at hadron colliders. Assuming
$C$, $P$ and SU(2) invariance, the LEP experiments have determined the
independent couplings to~\cite{leptgc}
\begin{eqnarray*}
g_1^Z & =  & \phantom{+}0.984^{+0.022}_{-0.019}, \\
\kappa_\gamma & = & \phantom{+}0.973^{+0.044}_{-0.045}, \\
\lambda_\gamma &  = & -0.028^{+0.020}_{-0.021}.  
\end{eqnarray*}

$W^+W^-$ production is sensitive to both the $WW\gamma$ and the $WWZ$
couplings. To measure these couplings independently, one has to consider
$W\gamma$ and $WZ$ production in hadronic collisions. Measurements of the
$WW\gamma$ couplings in $W\gamma$ production have been performed in
Run~I~\cite{Ellison:1998uy} and in Run~II~\cite{goshaw}. 
The D\O\ Collaboration recently presented the first direct
measurement of the $WWZ$ couplings from $WZ$ production. For 0.3~fb$^{-1}$,
and assuming $\Lambda=1.5$~TeV, they found that~\cite{Abazov:2005ys}
\begin{eqnarray*}
-0.48 & <\lambda_Z<0.48 \qquad {\rm for}~\kappa_Z & =
g_1^Z=1,\\ 
0.51 &<g_1^Z<1.66 \qquad {\rm for}~\lambda_Z & 
=\kappa_Z-1=0
\end{eqnarray*}
at 95\%~CL. Note that, at hadron colliders, TGC limits depend on the
form factor scale, $\Lambda$. 

Bounds on TGCs from hadron collider experiments scale roughly with
$(\int{\cal L}dt)^{1/4}$. One thus expects that the ultimate precision
which can 
be reached for the $WWV$ couplings at the Tevatron will be a factor~1.6
to~2.5 better than that obtained from current data, depending on the
final integrated luminosity. While the TGC bounds at the Tevatron only
mildly depend on the form factor scale, the dependence on $\Lambda$ is
much more pronounced at the LHC. In general, the $WWV$ couplings can be
measured with a precision of ${\cal O}(10^{-2}-10^{-3})$ at the
LHC~\cite{Haywood:1999qg}. 

\subsection{Theory of Di-boson Production}

\begin{figure}
\begin{center}
\resizebox{0.45\textwidth}{!}{
  \includegraphics{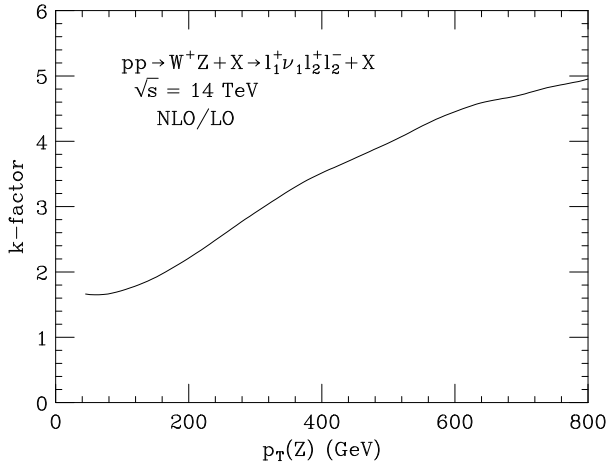}}
\caption[]{The ratio of the NLO to LO cross sections as a function of
$p_T(Z)$ for $W^+Z$ production at the LHC. }  
\label{fig:seven}       
\end{center}
\end{figure}

All di-boson production processes are known to NLO in
QCD~\cite{Ohnemus:1990za}. At the LHC QCD corrections to di-boson
production are large and increase with the $p_T$ of the vector
bosons. The ratio of the NLO to LO cross sections ($k$-factor) for
$W^+Z$ production at the LHC as a function of $p_T(Z)$ is shown in
Fig.~\ref{fig:seven}. Qualitatively similar results are obtained for the
other di-boson processes.
Since the QCD corrections give an effect which is qualitatively similar
to that of anomalous TGCs, it will be essential to take them into
account in any LHC di-boson analysis.

The EW radiative corrections to the di-boson production processes are also
known~\cite{Accomando:2001fn}. As in the case of single weak boson
production, they become significant at large energies, due to EW Sudakov
logarithms. For invariant masses in the TeV region they reduce the cross
section by typically $5-20\%$.

\section{Higgs Boson Physics}

The search for the SM Higgs boson is one of the main objectives of the
LHC. Over the last decade, enormous progress has been made in providing
accurate predictions for Higgs boson production and decays. In addition,
in the last few years, many studies of how well the
Higgs properties can be determined once this particle has been found
have been performed. 

The NLO QCD corrections to Higgs production via gluon fusion have been
calculated more than 10~years ago~\cite{Spira:1995rr}. They enhance the
$gg\to H$ cross section by a factor $1.5-2$. More recently,
several groups have calculated the NNLO QCD corrections to the total
$gg\to H$ cross section in the $m_t\to\infty$
limit~\cite{Harlander:2002wh}, showing that the perturbative series 
starts to converge at this order. A fully differential NNLO calculation for
$gg\to H\to\gamma\gamma$ also
exists~\cite{Anastasiou:2005qj}. Finally, the ${\cal O}(\alpha)$
corrections to Higgs production via gluon fusion have been
computed~\cite{Aglietti:2004ki}. They change the Higgs production cross
section by $5-8\%$ if $m_H=115-160$~GeV.

For $m_H<200$~GeV, production via vector boson fusion (VBF),
$qq'\to Hqq'$, is an important source for Higgs bosons. The QCD
corrections to $qq'\to Hqq'$ have been found to be
modest~\cite{Han:1992hr}. Associated production of Higgs bosons and top
quarks, $pp\to t\bar tH$, is a tool for measuring the top quark Yukawa
coupling. At LO, the $t\bar tH$ cross section strongly depends on the
factorization and renormalization scales. Once NLO QCD corrections are
taken into account, this dependence is greatly
reduced~\cite{Beenakker:2002nc}. 

While Higgs production is well under control theoretically, more
reliable calculations are still needed for several background
processes. In particular, calculations of the NLO QCD corrections are
needed for $t\bar tj$, $t\bar tb\bar b$ and EW $WWjj$ production.

Once a Higgs candidate has been found, one would like to determine how
the new particle couples to fermions, gauge bosons, and to
itself. Several studies have shown that, with mild theoretical
assumptions, the couplings of the new particle to fermions and gauge 
bosons can be measured with a precision of $10-30\%$ at the
LHC~\cite{Duhrssen:2004cv}. 

\begin{figure}
\begin{center}
\resizebox{0.45\textwidth}{!}{
  \includegraphics{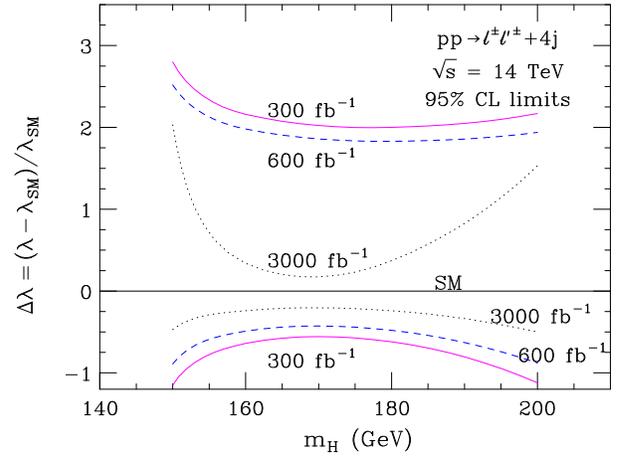}}
\caption[]{Limits achievable at 95\% CL for the normalized Higgs boson
self-coupling, $\Delta\lambda_{HHH}=(\lambda
-\lambda_{SM})/\lambda_{SM}$, at the LHC.} 
\label{fig:eight}       
\end{center}
\end{figure}

A measurement of the three Higgs boson self-coupling, $\lambda$, with a
similar precision is
considerably more difficult. In order to probe $\lambda$, one has
to study Higgs pair production, $gg\to HH$. For $m_H\geq 150$~GeV, the
$HH\to 4W\to\ell^\pm{\ell'}^\pm + 4$j channel offers the best
chances~\cite{Gianotti:2002xx,Baur:2002rb}. As shown in
Fig.~\ref{fig:eight}, with 300~fb$^{-1}$, it may
be possible to rule out a vanishing of $\lambda$ for
$m_H=150-200$~GeV, and measure the $HHH$ coupling with up to 20\%
accuracy at a SuperLHC with 3~ab$^{-1}$. 
For $m_H\leq 140$~GeV, $HH\to b\bar b\gamma\gamma$ is the most promising
final state. However, due to the tiny signal cross section in this
channel, a luminosity upgrade for the LHC is needed. Even with
3~ab$^{-1}$ one can only hope to achieve a precision of about 70\% for
$\lambda$~\cite{Baur:2003gp}.

While the signal to background ratio is of ${\cal O}(1)$ for $HH\to b\bar
b\gamma\gamma$, it is roughly $1/5-1/10$ for the $\ell^\pm{\ell'}^\pm +
4$j final state. The largest backgrounds contributing to $\ell^\pm{\ell'}^\pm +
4$j production originate from $WWWjj$, $t\bar tW$ and $t\bar tj$
production. The $t\bar tj$ background is particularly sensitive to the
acceptance cuts imposed, and thus tricky to estimate. More realistic
simulations for this background are needed. Furthermore, both
signal and background cross sections show a significant scale dependence
which could be reduced by full calculations of the NLO QCD corrections
to $gg\to HH$ (for finite $m_t$), and the background reactions. None of
these exist at the moment. 

\section{Summary}

Electroweak physics at hadron colliders is precision physics. Accurate
predictions are needed to fully utilize the potential of the Tevatron
and LHC for electroweak measurements. The theoretical predictions for
weak boson and Higgs boson production have become increasingly accurate
over the past few years. However, there is still much to do. In
particular a calculation which combines QCD and EW radiative corrections
for $W$ and $Z$ is needed, as well as calculations of the NLO QCD
corrections to a number of processes which contribute to the background
in Higgs production. 

\section*{Acknowledgments}

This research was supported by the National Science Foundation under
grant No.~PHY-0139953.

\end{document}